\documentclass[structabstract]{aa}  
\usepackage{graphicx}
\usepackage{natbib}
\bibpunct{(}{)}{;}{a}{}{,}
\usepackage{txfonts}
\begin{document}
   \title{Ionization driven molecular outflow in K3-50A}

   \author{P.D. Klaassen
          \inst{1}
          \and
          R. Galv\'an-Madrid\inst{2}
          \and
          T. Peters\inst{3}
          \and
          S.N. Longmore\inst{2}
          \and 
          M. Maercker\inst{2,4}
                    }

   \institute{Leiden Observatory,  PO Box 9513, 2300 RA, Leiden, The Netherlands
              \email{klaassen@strw.leidenuniv.nl}
         \and
             European Southern Observatory, Karl Schwarzschild Str 2, 85748 Garching, Germany
         \and
             Institut f\"{u}r Theoretische Physik, Universit\"{a}t Z\"{u}rich, Winterthurerstrasse 190, CH-8057 Z\"{u}rich, Switzerland
          \and
          Argelander Institut f\"{u}r Astronomie, Universit\"{a}t Bonn, Auf dem H\"{u}gel 71, 53121 Bonn, Germany
             }

   \date{}

 
  \abstract
  {Whether high mass stars continue to accrete material beyond the formation of an HII region is still an open question.  Ionized infall and outflow have been seen in some sources, but their ties to the surrounding molecular gas are not well constrained.}
   {We aim to quantify the ionized and molecular gas dynamics in a high mass star forming region (K3-50A) and their interaction.}
   {We present CARMA observations of the 3mm continuum, HCO$^+$ and H41$\alpha$ emission, and VLA continuum observations at 23 GHz and 14.7 GHz to quantify the gas and its dynamics in K3-50A. }
   {We find large scale dynamics consistent with previous observations.  On small scales, we find evidence for interaction between the ionized and molecular gas which suggests the ionized outflow is entraining the molecular one. This is the first time such an outflow entrained by photo ionized gas  has been observed.}
   {Accretion may be ongoing in K3-50A because an ionized bipolar outflow is still being powered, which is in turn entraining part of the surrounding molecular gas. This outflow scenario is similar to that predicted by ionization feedback models.}


   \maketitle
%

\section{Introduction}

The processes involved in the formation of high mass stars are generally said to be similar to, but scaled up from the processes responsible for forming low mass stars.  Many models even suggest that high mass stars can continue to accrete material once they are more than 8 M$_\odot$ by either puffing up \citep[e.g.][]{Hosokawa09} which delays the onset of an HII region, or via funneled accretion flows \citep[e.g.][]{Keto02,Keto03} despite the onset of an HII region. Simulations further indicate that partial shielding of the ionizing radiation by dense filaments in the accretion flow is the key effect that allows massive stars to grow after an HII region has formed \citep{Peters:2010bw}.

 There is growing observational evidence that molecular gas is continuing to fall onto HII regions \citep[e.g.][]{K11}, as well as, in some cases, continued ionized infall\citep[e.g.][]{KW06,Beltran06}. Through understanding the dynamics in high mass star forming regions, we can make direct comparisons between models and observations.
 
 Molecular gas dynamics are detected using various species depending on the ambient density, temperature, etc, while  radio recombination lines (RRLs) can be used to determine the dynamics of the ionized gas.    \citet{Mottram11} suggest that ongoing ionized accretion has a measurable lifetime only in regions with bolometric luminosities of greater than about 10$^5$ L$_\odot$. The high mass star forming region K3-50A, with a bolometric luminosity of $2\times10^6$ L$_\odot$ \citep{Kurtz94} is thus an ideal candidate for tracing these latest stages of the formation of truly massive stars.

K3-50A, at a distance of 8.7 kpc \citep[i.e.][]{okamoto03}, has been studied at wavelengths ranging from radio to the infrared. In previous observations, the 14.7 GHz continuum was shown to have a bipolar morphology, and was modeled as a bipolar ionized outflow which was in agreement with the RRL (H76$\alpha$) observations of the source \citep{depree94}. \citet{Howard97} showed the HCO$^+$ J=1-0 emission for this source is elongated perpendicular to the outflow, and gave the rest velocity of the source as -23.7 km s$^{-1}$.  They estimated this envelope molecular gas mass to be 2600 M$_\odot$, and  suggested, based on fitting radiative transfer models, that the HCO$^+$ is tracing a large scale (0.75 pc) rotating toroidal structure. Their modeling suggests an inclination angle of 30$^\circ$ to the line of sight, and they note that some of the small scale HCO$^+$ surrounding the core of the HII region appears to have a larger velocity dispersion which they attribute to interaction with the ionized outflow.   There is no evidence for a large scale molecular outflow in this region.  \citeauthor{Howard97} did however detect SiO on small scales towards the center of this region (with a $\sim 2.5''$ beam). \citet{okamoto03} identified three 16 M$_\odot$ stars within the HII region via  infrared observations, which they suggest may be powering the HII region.

Here, we present CARMA and VLA observations of K3-50A at 90 and 23 GHz.  The 90 GHz observations are at higher resolution than the \citet{Howard97} observations, and show evidence for a small scale molecular outflow being driven by the ionized one.  In Section \ref{sec:observations} we present our observations, and in Section \ref{sec:results} we present our results and interpretation of the data both on small and large scales.  In Section \ref{sec:discussion} we discuss this ionization driven outflow in terms of model predictions, and in Section \ref{sec:conclusions} we conclude.

\section{Observations}
\label{sec:observations}

CARMA observations of K3-50A were taken in the C configuration on November 9 and 10, 2008.  Narrowband observations were taken in all spectral windows. This required using  simultaneous wideband and hybrid (combined wide and narrow band) observations of the calibrators to achieve the required signal to noise ratio for calibration\citep[i.e.][]{wei08}.   The on-site NOISE source (a transmitter on the CARMA site) was used to help  bandpass and gain calibration as its signal is strong, stable, and well characterized. Bandpass calibrations were first carried out on the local NOISE source in the wideband spectral windows, and these calibrations were applied to the wideband observations of the bandpass calibrator. These spectral windows were then bandpass calibrated using the astronomical bandpass calibrator (as listed in Table \ref{tab:calibrators}), and this bandpass calibration was applied to the narrowband spectral windows. The same process was then repeated for the narrowband windows on the bandpass calibrator and further applied to K3-50A.The flux calibrator, MWC349, was assumed to have a flux of 1.29 Jy at 90 GHz. The calibrators and beam statistics can be found in Table \ref{tab:calibrators}. Over the two nights, there was a total on source time of 3.8 hr. The CARMA data were calibrated using MIRIAD.

Four of the six spectral windows were combined to determine the continuum emission, and the other two were used for line emission.  H41$\alpha$ was observed in the upper sideband at 92.03 GHz, while HCO$^+$ (J=1-0) was observed in the lower sideband at 89.188 GHz. The bandwidth in each narrowband spectral window was 31 MHz, with a spectral resolution of 0.488 MHz (1.7 km s$^{-1}$). 

For the continuum emission, the observed spectral windows were concatenated together in the UV plane, and then imaged.  Self-calibration was done in both phase and amplitude.  For H41$\alpha$, imaging and self calibration were done in a similar fashion for the single spectral window. The two nights of  HCO$^+$ observations were best self calibrated separately, and then combined while inverting the visibilities to the image plane. Attempts to self calibrate the bright and large scale HCO$^+$ emission using the small scale continuum did not work.  However, using the strong ling emission itself to self calibrate was successful. The resulting map is consistent with the lower resolution map of \citet{Howard97}. The CARMA continuum, ionized, and molecular gas datasets were inverted using robust weightings of 0, -2 and 2 (respectively) which yielded the best signal to noise ratio for each component, each of which emits on different size scales. In the continuum,  HCO$^+$, and H41$\alpha$ maps, we achieved single channel noise levels of 6.8,  55, and 56 mJy/beam (respectively).

\begin{figure*}[htb!]
\centering
\includegraphics[ bb = 18 200 594 600,width=0.9\textwidth]{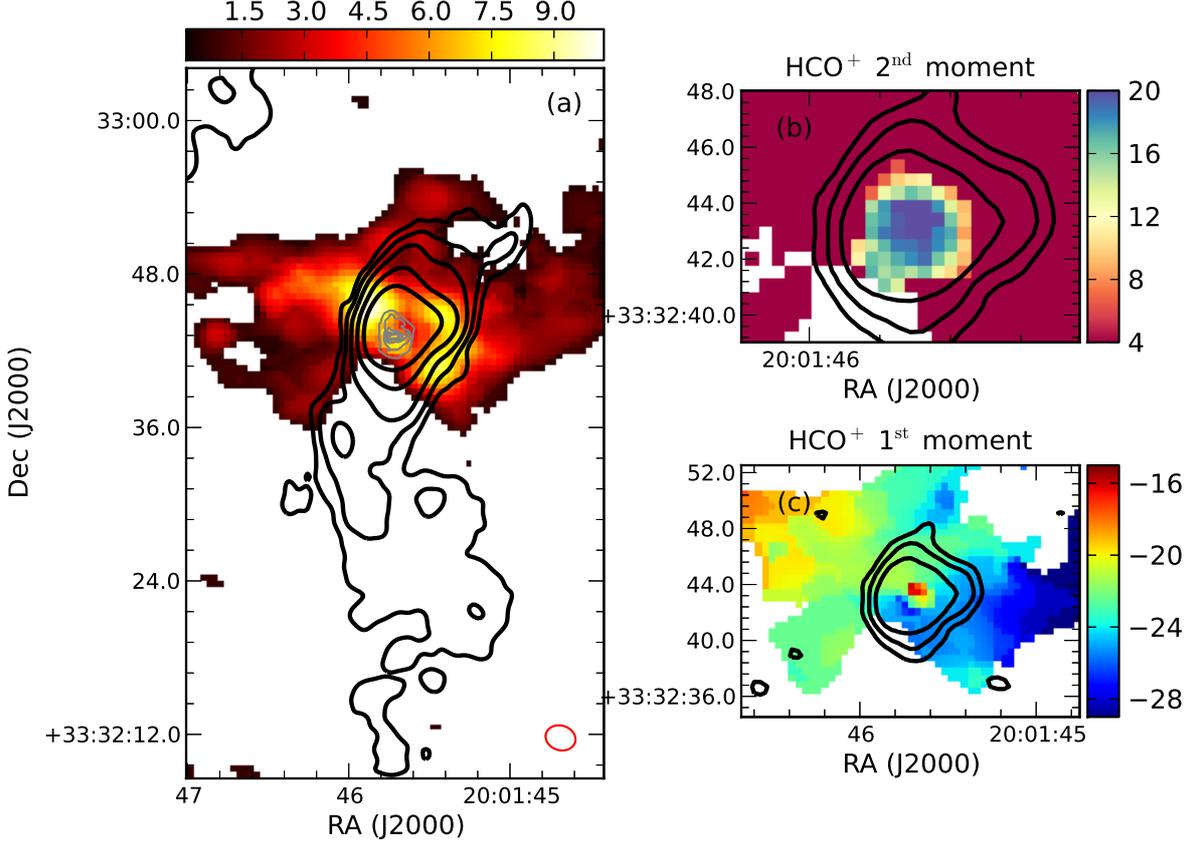}
\caption{HCO$^+$ in K3-50A.  (a) HCO$^+$ (J=1-0) integrated intensity with VLA 14.7 and 23GHz continuum contours in black and grey. The contours are  3, 10, 30, 100 and 300 times the rms noise levels given in Table \ref{tab:calibrators}.  The color scale has units of Jy beam$^{-1}$ km s$^{-1}$, and the beam is shown in the bottom right corner.  (b) HCO$^+$ velocity dispersion map. The color scale has units of km s$^{-1}$ and shows a sharp change in velocity dispersion at the edges of the HII region. The black contours here show the 90 GHz continuum, and are 3, 10 and 30 times the 90GHz continuum rms noise (6.8 mJy beam$^{-1}$). (c) The HCO$^+$ intensity weighted velocity map. The color scale units are km s$^{-1}$, and the contours are the same as in panel (b). }
\label{fig:hco_moments}
\end{figure*}

VLA B configuration observations of the 23 GHz continuum were taken on 16 and 17 September 2006.  The data were reduced in CASA, and both JVLA outfitted antennas were removed from the dataset.  The calibrators and beam statistics are shown in Table \ref{tab:calibrators}. The source was self-calibrated and imaged using natural weighting.   The rms noise in this map is 0.96 mJy/beam.
The archival VLA 14.7 GHz continuum observations presented in \citet{depree94} were reprocessed in AIPS.  The continuum emission was derived from the 18 line free channels of the 31 channel bandpass. The final continuum flux was within 2\% of that given in \citet{depree94}. The rms noise in this map was 0.64 mJy/beam

\begin{table}
\caption{VLA (23 GHz) and CARMA Calibrators and Beam Properties}
\begin{tabular}{lrrrr}
\hline \hline
Telescope &  \multicolumn{3}{c}{Calibrators} & \multicolumn{1}{c}{Beam}\\
\cline{2-4}
	 & Bandpass & Phase & Flux &HPBW($''$), PA \\	\hline
VLA		& 2253+161	& 2025$+$337 & 3C286 & 0.31$\times$0.28, 76$^\circ$\\
CARMA 	& 2148+069  & 2015+372 & MWC349 & 1.80$\times$1.50, 81$^\circ$\\
\hline
\end{tabular}
{ The continuum rms noise levels are 0.96 and 6.8 mJy/beam for the VLA (23 GHz) and CARMA (90 GHz) observations respectively.}
 \label{tab:calibrators}
\end{table}
\vspace{-10pt}


\section{Results}
\label{sec:results}

Below, we present an analysis of our CARMA and VLA observations of K3-50A.  We first discuss the radio continuum properties and how we derived the ionized gas mass from it.    We then go on to describe how our H41$\alpha$ observations support the previous claims of a bipolar ionized outflow in this region and how, on small scales, this ionized outflow is affecting and entraining the molecular gas around it.  We finish by briefly describing the large scale HCO$^+$.  In Figure \ref{fig:hco_moments} we show the zeroth, first and second moment maps of the molecular gas observed with CARMA.  In the left panel we show the integrated intensity of the HCO$^+$ J=1-0 emission along with the 14.7 and 23 GHz continuum emission from the HII region.  In the right two panels we show the velocity dispersion or second moment (top) and intensity weighted velocity or first moment (bottom) of the HCO$^+$ over plotted with 90 GHz continuum emission contours.

\subsection{Continuum emission and ionized gas mass}
\label{sec:cont}

The 14.7 and 23 GHz continuum emission observed at the VLA are presented as black and grey contours (respectively) in panel (a) of Figure \ref{fig:hco_moments}.  The detailed structure of the 23 GHz continuum emission is also shown by grey contours in the H41$\alpha$ first moment map of Figure \ref{fig:outflow}.  The integrated and peak flux densities of the continuum emission are given in Table \ref{tab:continuum}, along with the deconvolved full width at half maximum sizes of the emitting regions.   In addition to the continuum values derived from our VLA observations, the 90 GHz continuum observations with CARMA are also shown by black contours in panels (b) and (c) of Figure \ref{fig:hco_moments} and the first moment maps of Figure \ref{fig:outflow}.The 90 GHz continuum properties are listed in Table \ref{tab:continuum}. \citet{depree94} determined the emission measure (EM, 5.8$\times10^8$ cm$^{-6}$ pc), electron temperature (T$_e$, 8\,000 K) and electron column density ($n_e$, 7.9$\times10^4$ cm$^{-3}$) for K3-50A.  We used the emission measure and temperature to determine the continuum optical depth at 14.7 and 23 GHz to be 0.67 and 0.26 (respectively) using equation 10.36 of \citet{Tools5}. The continuum optical depth at both 14.7 and 23 GHz is less than one, however it is not completely optically thin ($\tau <0.1$). We will however, for the sake of simplicity, assume the 23 GHz emission ($\tau=0.26$) is optically thin for calculating the ionized gas mass.

The mass of ionized gas in the HII region can be determined using the relations derived in \citet{mezger67} with 5 GHz continuum, assuming the continuum is optically thin.  \citet{GalvanMadrid08} extrapolated this equation to other frequencies, and we use their equation to derive the mass of ionized gas in the HII region as traced by the 14.7 and 23 GHz continuum observations. We used a source diameter of 5$''$ (0.21 pc) for these calculations as that encloses all of the 23 GHz emission, and roughly corresponds to the 20\% power annulus of the 14.7 GHz continuum, and the characteristic shell described in \citet{depree94}. The integrated fluxes listed in Table \ref{tab:continuum} were used for this calculation.  We find the two datasets give ionized gas masses of 8 and 7.1 M$_\odot$. Note that  the emission at neither frequency is completely optically thin, but as the optical depth at 23 GHz is lower by a factor of three, we use the ionized gas mass derived from that emission (8 M$_\odot$).  We did not calculate the ionized gas mass from the 90 GHz continuum because there is potentially a thermal dust component at that frequency, and the SED is not well enough constrained on small scales to distinguish between the two components.

\begin{table}[hbt]
\caption{Measured Continuum Parameters}
\begin{tabular}{crrrrr@{$\times$}rr}
\hline
\hline
Freq.& \multicolumn{2}{c}{Flux Density}&Noise & Mass & \multicolumn{3}{c}{Deconvolved FWHM}\\
(GHz) &Integrated & Peak & &(M$_\odot$)&\multicolumn{2}{c}{($''\times ''$)} & PA\\
\hline
90 & 8.5 & 3.2 &6.84& -$^{a}$& 2.03& 1.89  &-34.\\
23 & 7.1 & 0.3 &0.96& 8&1.92 & 1.8 & -.05\\
14.7 & 4.8 & 2.1 &0.64& 7.1& \multicolumn{3}{c}{bipolar \& extended}\\
\hline
\end{tabular}
{ Integrated and peak flux densities (in units of Jy and Jy beam$^{-1}$ respectively), rms noise level (in units of mJy beam$^{-1}$), ionized gas mass, and deconvolved continuum size}\\
$^{a}${Ionized gas mass not calculated at 90 GHz due to multiple emission components contributing to the integrated flux (i.e. free-free and thermal dust).}
\label{tab:continuum}
\end{table}

\subsection{H41$\alpha$ tracing the ionized outflow}
\label{sec:ionized}

\begin{figure}[hbt]
\centering
\includegraphics[width=\columnwidth]{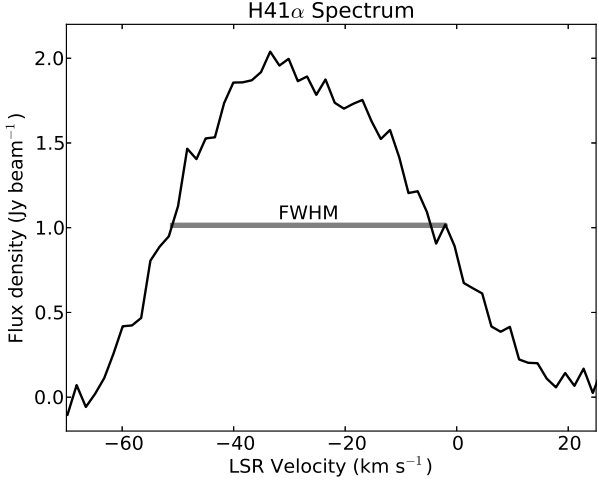}
\caption{ H41$\alpha$ spectrum integrated over the 90GHz continuum region. We highlight the line FWHM with a grey line. }
\label{fig:H41spec}
\end{figure}

Previous observations of H76$\alpha$ showed a velocity gradient within the HII region in K3-50A.  The direction of the velocity gradient was consistent with the direction of the bipolar nebula leading \citet{depree94} to suggest that there is an ionized outflow coming from this HII region.   
With significant line broadening due to pressure, dynamical properties in the ionized gas are hard to quantify at these low frequencies.  However, pressure broadening varies with $n$ (the electron orbital number) to the power of 7.4 \citep[e.g.][]{BS72}. For H41$\alpha$, the line width contribution from pressure broadening is much lower than that in the H76$\alpha$ transition (by a factor of a few hundred), and falls well below our channel width.   This makes quantifying the ionized gas dynamics at (sub-)mm wavelengths more straightforward, especially when, at higher frequencies, the ratio of the RRL  to continuum emission is also much more favourable for detecting line emission.  For our H41$\alpha$ line, we use a FWHM of 50 km s$^{-1}$ which is smaller than the 80 km s$^{-1}$ FWHM of the H76$\alpha$ line as measured from panel (b) of Figure 7 in \citet{depree94}.  Since the pressure broadening of the line should be much lower in H41$\alpha$ than H76$\alpha$, the only slight decrease in line width (from 80 to 50 km s$^{-1}$ suggest unresolved dynamics in the gas. Using equation 14.29 of \citet{Tools5}, we calculated that at 92 GHz, the line to continuum ratio is $T_L/T_c=0.6$ for a line width of 50 km s$^{-1}$ (see Figure \ref{fig:H41spec}). Because of the effects of less pressure broadening and relatively brighter lines at 90 GHz combined with our higher velocity resolution \citep[1.7 km s$^{-1}$ compared to the 8 km s$^{-1}$ of ][]{depree94}, we are able to  better constrain the velocity gradients in the ionized gas of K3-50A.

\begin{figure*}
\centering
\includegraphics[width=\textwidth]{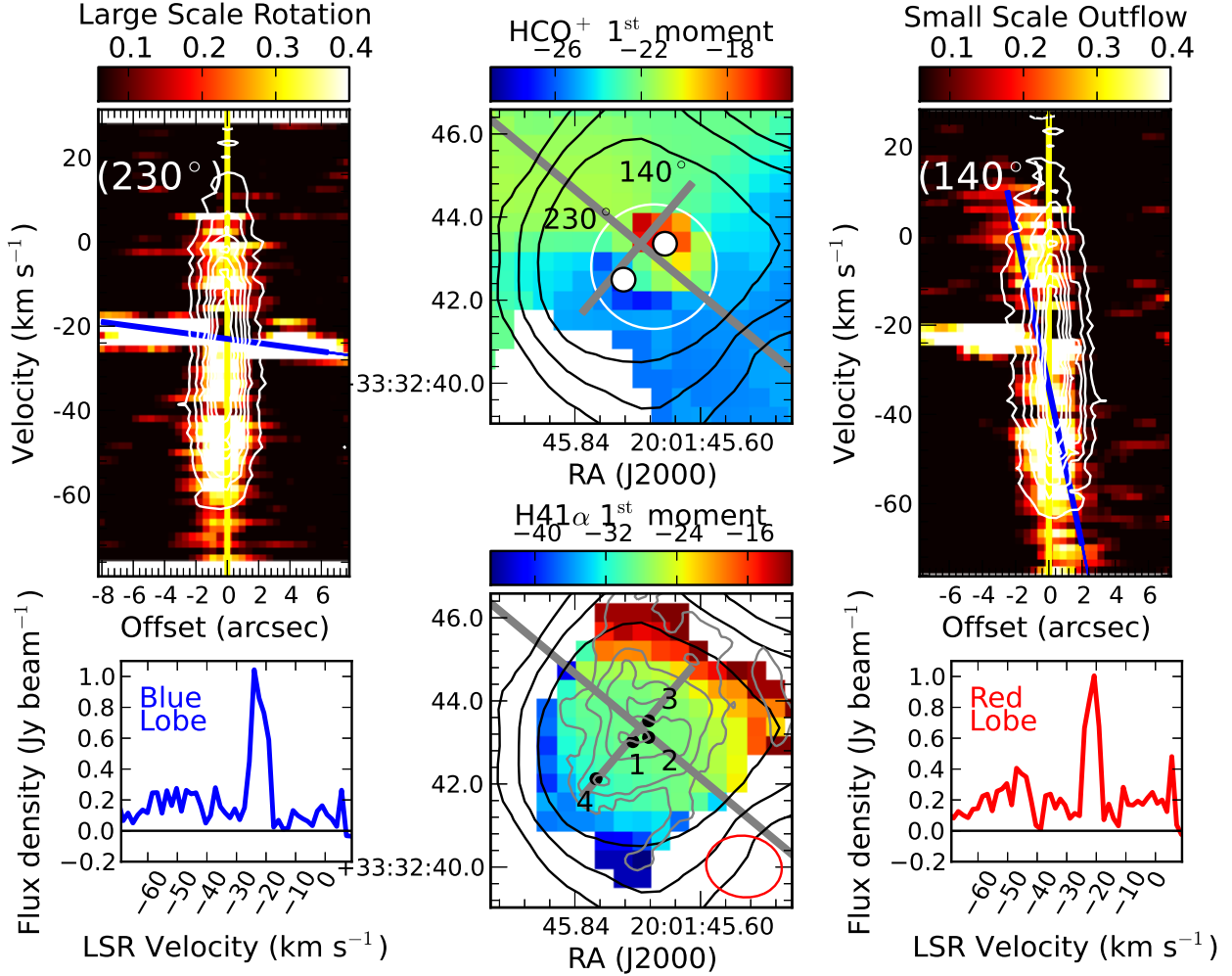}
\caption{{\bf First Moment Maps:} The middle two panels show the first moment maps of the HCO$^+$ (top) and H41$\alpha$ emission towards the center of K3-50A. with their spreads in velocities noted at the top of each panel. The two grey lines show the cuts used for the two PV diagrams shown to the left and right. The labelled black dots in the H41$\alpha$ map show the positions of the four IR sources in \citet{okamoto03} which are likely high mass protostars. The black contours show the 90 GHz continuum emission as in panels (b) and (c) of Figure \ref{fig:hco_moments}, while the grey contours show the 23 GHz continuum (at 10, 30, 100 and 300$\sigma$). The red circle at the bottom right shows the synthesized CARMA beam. The white dots on the HCO$^+$ map show the positions used for the two spectra displayed here.  {\bf PV Diagrams:} The two PV diagrams show the  ionized (contours) and molecular (colourscale) emission along the outflow direction (at a position angle of 140$^\circ$) as well as the emission perpendicular to it (at a position angle of 230$^\circ$). The yellow line in each panel shows the source center, and it is clear that there is a velocity gradient in the high velocity dispersion gas at 140$^\circ$. The large scale rotation at a position of 230$^\circ$ manifests in the low velocity dispersion emission. These velocity gradients are highlighted with blue lines.}
\label{fig:outflow}
\end{figure*}

In Figure \ref{fig:outflow} we show the dynamics of the small scale ionized and molecular gas. The middle panels show the first moment maps of the H41$\alpha$ and HCO$^+$ emission.  Discussion of the molecular gas dynamics on these size scales follows below in Section \ref{sec:outflow}.  The two grey lines over plotted on the first moment maps show the cuts used to create the two position-velocity (PV) diagrams in Figure \ref{fig:outflow}.  One PV diagram is taken along the direction of the H41$\alpha$ velocity gradient (labelled with its position angle of 140$^\circ$, and presented in the same figure) and one perpendicular to it (labelled with its position angle of 230$^\circ$, and presented to the left in the figure). The H41$\alpha$ emission is shown as contours in the PV diagrams.

  In the first moment map there is a clear velocity gradient in the H41$\alpha$ seen along the same direction first presented in \citet{depree94}.   We find a velocity gradient of

\begin{eqnarray}  
  \frac{\Delta V}{\Delta R} &=& \frac{ 25 {\rm km\, s}^{-1}}{0.21{\rm pc}} =  120\pm13 {\rm km\, s}^{-1} {\rm pc}^{-1} 
  \label{eqn:gradient}
  \end{eqnarray}
 
  where the uncertainties are derived from 0.2 of a beam, and half of the velocity resolution of the observations and are propagated  in quadrature. The position angle of the velocity gradient is $\sim 140^\circ$ and is shown by one of the grey lines in Figure \ref{fig:outflow}.  Quantitatively, it is unlikely that the gradient is due to rotation.  If we were to assume it were due to Keplerian rotation would indicate an enclosed mass of 3$\times10^4$ M$_\odot$, more than 10 times the mass of the surrounding parsec scale envelope as traced by HCO$^+$ (see Section \ref{sec:hco_infall}).   In the first moment map of H41$\alpha$ we show the velocity shift across the source ($\Delta V$ = 25 km s$^{-1}$). Using a characteristic velocity of half of the total velocity shift (12.5 km s$^{-1}$), and the mass of ionized gas calculated from the 23 GHz continuum emission in Section \ref{sec:cont} (8 M$_\odot$), we determine an outflow momentum of 100 M$_\odot$ km s$^{-1}$.  Using the size in Equation \ref{eqn:gradient}, we determine an outflow force  averaged over the area presented in the bottom middle panel of Figure \ref{fig:outflow} using  
  
  \begin{eqnarray}
  F = \frac{MV^2}{R} &=& \frac{8{\rm \,M}_\odot * (12.5 {\rm \,km\, s}^{-1})^2}{0.21{\rm \,pc}}\label{eqn:force}\\
  &=&6.0\times10^{-3} {\rm M\,}_\odot {\rm \,km\, s}^{-1} {\rm \,yr}^{-1}.
  \end{eqnarray}

    We cannot use the method for determining the momentum generally used for molecular gas (of summing the momenta in each channel) because our ionized gas mass was derived from the continuum. We also note that there are studies of the differences between various methods of determining outflow momentum and force, and that for low mass sources \citep[e.g.][and van der Marel et al. 2013]{Cabrit90,Cabrit92,Downes07} these studies show that the differences between momentum calculations using a characteristic velocity, as was done here, or summing the momenta over each channel, only vary by a factor of a few. A likely larger source of uncertainty in our momentum estimate comes from the distance uncertainty, as the mass varies as D$^{2.5}$.

\subsection{Small Scale HCO$^+$: tracing an entrained outflow}
\label{sec:outflow}

As shown in panel (b) of Figure \ref{fig:hco_moments}, on small scales, the velocity dispersion of the HCO$^+$ increases dramatically, from 4 km s$^{-1}$ to upwards of 20 km s$^{-1}$.  This increase in velocity dispersion is unlikely to be due to thermal broadening (which would require a temperature of 10$^5$ K, too hot for molecular gas to be present), and is likely due to the molecular gas being stirred up by something. We discuss below how it is the ionized gas agitating and entraining the molecular gas.  

In the HCO$^+$ first moment map of Figure \ref{fig:outflow}, we zoom in on the region of high velocity dispersion, and highlight it with an empty white circle. The two white points on the plot show the positions used for the spectra plotted in the same figure.  The colourscales in the PV diagrams show the HCO$^+$ emission along the two grey lines shown on the first moment map.  The high velocity dispersion gas shows its large velocity ranges in both diagrams. In the cut taken along the ionized outflow direction (labelled as 140$^\circ$), there is a velocity gradient in the molecular gas which we highlight with a blue line.  This NW-SE velocity gradient extends from -28 km s$^{-1}$ to -16 km s$^{-1}$, and is approximately two synthesized beams across. While the velocity shift is smaller than that in the ionized gas, the velocity gradient \mbox{($\Delta V/\Delta R$ =  12 km s$^{-1}$ / 0.084 pc = 142$\pm$23 km s$^{-1}$ pc$^{-1}$)} is the same as that derived for H41$\alpha$ above.  

A diagonal line in a PV diagram is indicative of a velocity gradient due to either rotation or outflow.  We reject the rotation hypothesis because if the velocity gradient were due to Keplerian rotation, the data would suggest a mass of 15\,000 M$_\odot$ within 0.2 pc.  This is much larger than the 2\,600 M$_\odot$ previously calculated for the large scale envelope of K3-50A \citep{Howard97}. Thus, the velocity gradient must be due to outflow motions.

Under the assumption that the high velocity dispersion gas is outflowing, we calculated the outflow properties shown in Table \ref{tab:comparison} from the mass and velocity of the outflowing material. Using the same assumptions as described below in Section \ref{sec:hco_infall} to calculate the molecular gas mass from the HCO$^+$ emission within a diameter of 2.5$''$ (where the velocity dispersion is high in panel (b) of Figure \ref{fig:hco_moments}), we determine an outflow mass of $M_{\rm out}$=4.6 M$_\odot$.  The characteristic velocity (half the total velocity shift) of this outflowing molecular gas is $V_{\rm out}$=6 km s$^{-1}$. From these mass and velocity estimates, we calculated the total molecular outflow momentum:

\begin{equation}
P=M_{\rm out}V_{\rm out}= 28 {\rm\, M}_\odot {\rm \,km\,s}^{-1}.
\end{equation}

\noindent This momentum derivation method, which is consistent with the way the ionized momentum was calculated, yields similar results to summing the momentum in each channel (21 M$_\odot$ km s$^{-1}$).\,   Additionally, we calculated the total outflow mechanical energy ($E=0.5 M_{\rm out}V_{\rm out}^2=1.6\times10^{45}$ erg) and outflow force (3.2$\times10^{-3}$ M$_\odot$ km s$^{-1}$ yr$^{-1}$, see equation \ref{eqn:force}), where we used a radius of 1.25$''$ ($R$=0.0527pc). All of these values are derived assuming the HCO$^+$ is optically thin. We cannot constrain the opacity of the HCO$^+$ because we were unable to observe an optically thin tracer in our observing setup.
\subsection{Large scale HCO$^+$: tracing the rotating envelope}
\label{sec:hco_infall}

The large scale HCO$^+$ around the HII region in K3-50A appears to be undergoing bulk dynamical motion as shown by the velocity gradient in panel (c) of Figure \ref{fig:hco_moments}, and the left PV diagram of Figure \ref{fig:outflow}, where it is highlighted by a blue line on 16$''$ scales. Our observations show coherent flows over the 1.2 pc extent of the HCO$^+$ emission. These large scale structures are consistent with those of the  HCO$^+$ J=4-3 observed with the JCMT (at 15$''$ resolution) by \citet{KW08}.  The BIMA observations of \citet{Howard97}, also in HCO$^+$ J=1-0, are well represented by a rotating torus of gas with an outer radius of 0.75 pc.   Using an HCO$^+$ abundance of 3.9$\times10^{-8}$ with respect to H$_2$ and an ambient temperature of 45 K previously determined for this source \citep{Howard97},  and assuming optically thick emission ($\tau=1$), we derive a gas mass of 2175 M$_\odot$.   This value is consistent with the lower resolution observations of \citet{Howard97}.

We determined characteristic velocities for the HCO$^+$ lobes to the east and west of the ionized outflow from first moment maps averaged over each lobe.  The velocity was multiplied by the mass of the lobe to determine its bulk momentum. Summing the momenta over the two lobes, we get a total cloud momentum of $\sim$7\,000 M$_\odot$ km s$^{-1}$.  
The combined mass and momentum, as traced by our interferometer observations, suggest that this gas reservoir is too large to be the result of an outflow \citep[c.f.][]{Wu04}.

\section{Discussion}
\label{sec:discussion}

As shown in \citet{depree94}, and in Figure \ref{fig:hco_moments}, K3-50A shows a large scale ionized outflow.  We have presented calculations for the velocity gradient (120 km s$^{-1}$ pc$^{-1}$) and momentum (100 M$_\odot$ km s$^{-1}$) in this ionized outflow. On small ($\sim$ few arcsec) scales, the ionized outflow is interacting with its surrounding molecular gas. This interaction has increased the HCO$^+$ velocity dispersion on these size scales (to 20 km s$^{-1}$), and has entrained some of this molecular gas into a small scale outflow.  This molecular outflow has a velocity gradient of 142 km s$^{-1}$ pc$^{-1}$ and a lower limit to its momentum of 28 M$_\odot$ km s$^{-1}$. The spatial and dynamical coincidence of the ionized and molecular outflows indicates a common origin for the two outflows.  That the momentum and force are lower in the molecular outflow suggests that the ionized outflow is entraining the molecular one, and only transferring a fraction of its momentum to the molecular gas.  This outflow entrainment mechanism is very different from the canonical outflow scenario in which the magnetic fields are responsible for generating an outflow, but is consistent with processes identified in simulations with ionization feedback \citep{Peters:2010bw} and is described in detail in  \citet{with_thomas_12}.

 In these ionization-driven outflows, the thermal pressure of the ionized gas ejects fountains of molecular material perpendicular to the accretion disk in which the massive star is embedded. During this driving phase, the outflow consists of a dense molecular shell with an ionized interior. However, since H II regions flicker on a short timescale as long as the powering massive star is still accreting \citep{Roberto_Thomas11}, the outflow interior does not remain ionized. When the ionizing radiation is shielded by dense filaments in the accretion flow, the ionized interior of the fountain recombines on a timescale of ~10-1000 yr, and the outflow is molecular with only a small or no visible H II region at the footpoint. These outflows were shown in simulated CO observations with ALMA in \citep{with_thomas_12}.

Models of outflows driven by ionization feedback show that the resultant molecular outflows are much smaller, in terms of mass, size, and energetics than those generally observed in high mass star forming regions.  They are small because they are not continuously driven, and their energetics are much lower for the same reasons.  Table \ref{tab:comparison} shows a comparison between the small scale outflow properties we have derived using HCO$^+$ in K3-50A and those seen in other high mass star forming regions of comparable bolometric luminosities ($>10^6$ L$_\odot$).  The outflow from K3-50A is orders of magnitude smaller than those seen in other regions, which suggests it is not a classically driven outflow. \citet{with_thomas_12} found no evidence that the outflow mass or energetics scale with the stellar luminosity.

The derived outflow properties are only a factor of a few higher than those predicted in \citet{with_thomas_12}.  It should be noted that their simulations were not attempting to explain regions quite as massive as K3-50A.  Their initial envelope mass was only 1\,000 M$_\odot$, which is much lower than the current envelope mass of 2\,200-2\,600 M$_\odot$ quoted in this work and \citet{Howard97}, from which stars have already formed.

Another characteristic of these models is that the protostar(s) must still be accreting, and that the accretion flow should be perpendicular to the outflow axis.  This is likely the case for K3-50A although we cannot constrain this with our observations. The Br$\gamma$ observations of \citet{Blum2009} show wide line widths, and red and blue shifted profiles (see their Figure 7) which are perpendicular to the outflow velocity gradients seen in our Figure \ref{fig:outflow}. These velocity shifts may be indications of accretion as Br$\gamma$ is generally used as an accretion tracer in lower mass star forming regions \citep[e.g.][]{Muzerolle98}.

\begin{table}
\caption{ K3-50A outflow properties compared to other sources}
\begin{tabular}{crlrrr}
\hline \hline
Source & L$_{\rm Bol}$ & SiO & \multicolumn{3}{c}{Molecular Outflow Properties}\\
	&			& detection & Mass & Momentum& Energy\\
	\multicolumn{2}{r}{ ($10^6$ L$_\odot$)} & (Y/N) & (M$_\odot$) & (M$_\odot$ km s$^{-1}$) & ($10^{45}$ erg)\\
	\hline
K3-50A & 2 & Y$^{a,b,d}$ & 4.6 & 28 & 1.65\\
\hline
G10.60$-$0.40 & 1.1$^c$ & Y$^{d}$ & 90$^c$ & 670$^c$ & 50$^c$\\
G10.47+0.03 & 1.4$^c$ & Y$^{a,d}$& 150$^c$ & 1110$^c$ & 81$^c$\\
G48.61+0.02 & 1.3$^c$ & Y$^{e}$& 590$^c$ & 2550$^c$ & 118$^c$\\
G19.61$-$0.23 & 1.7$^c$ & Y$^{d}$& 200$^c$ & 1740$^c$ & 150$^c$\\
\hline
\end{tabular}
{Comparison with outflows from other high luminosity ($>10^6$ L$_\odot$) star forming regions. Note that the outflow properties for the comparison sources also assume optically thin emission.}
\tablebib{(a) \citet{KW07}, (b) \citet{Howard97}, (c) \citet{LS09}, (d) \citet{KW08}, (e) \citet{K12}}
\label{tab:comparison}
\end{table}

\section{Conclusions}
\label{sec:conclusions}

We have quantified the small scale ionized and molecular gas dynamics in K3-50A, and show that we are likely seeing evidence for an ionization driven molecular outflow. The small scale molecular outflow observed for the first time in this source is orders of magnitude smaller than the more traditionally powered outflows seen from sources of comparable luminosity. This, combined with the the two outflows matching in size, scale and direction, suggests a different powering source for the molecular outflow, which we interpret as being the ionized outflow.  The tandem observations of the ionized and molecular gas dynamics have enabled us to quantitatively argue for an ionization driven molecular outflow. We would not have been able to do this only with observations of the ionized or molecular gas components; we required both.

\bibliographystyle{aa}

\begin{thebibliography}{29}
\expandafter\ifx\csname natexlab\endcsname\relax\def\natexlab#1{#1}\fi

\bibitem[{Beltran {et~al.}(2006)Beltran, Cesaroni, Codella, Testi, Furuya, \&
  Olmi}]{Beltran06}
Beltran, M.~T., Cesaroni, R., Codella, C., {et~al.} 2006, \nat, 443, 427

\bibitem[{Blum \& McGregor(2009)}]{Blum2009}
Blum, R.~D. \& McGregor, P.~J. 2009, The Astronomical Journal, 138, 489

\bibitem[{Brocklehurst \& Seaton(1972)}]{BS72}
Brocklehurst, M. \& Seaton, M.~J. 1972, Monthly Notices of the Royal
  Astronomical Society, 157, 179

\bibitem[{Cabrit \& Bertout(1990)}]{Cabrit90}
Cabrit, S. \& Bertout, C. 1990, Astrophysical Journal, 348, 530

\bibitem[{Cabrit \& Bertout(1992)}]{Cabrit92}
Cabrit, S. \& Bertout, C. 1992, Astronomy and Astrophysics (ISSN 0004-6361),
  261, 274

\bibitem[{De~Pree {et~al.}(1994)De~Pree, Goss, Palmer, \& Rubin}]{depree94}
De~Pree, C.~G., Goss, W.~M., Palmer, P., \& Rubin, R.~H. 1994, The
  Astrophysical Journal, 428, 670

\bibitem[{Downes \& Cabrit(2007)}]{Downes07}
Downes, T.~P. \& Cabrit, S. 2007, Astronomy and Astrophysics, 471, 873

\bibitem[{Galv{\'a}n-Madrid {et~al.}(2011)Galv{\'a}n-Madrid, Peters, Keto,
  Mac~Low, Banerjee, \& Klessen}]{Roberto_Thomas11}
Galv{\'a}n-Madrid, R., Peters, T., Keto, E.~R., {et~al.} 2011, Monthly Notices
  of the Royal Astronomical Society, 416, 1033

\bibitem[{Galv{\'a}n-Madrid {et~al.}(2008)Galv{\'a}n-Madrid, Rodr{\'\i}guez,
  Ho, \& Keto}]{GalvanMadrid08}
Galv{\'a}n-Madrid, R., Rodr{\'\i}guez, L.~F., Ho, P. T.~P., \& Keto, E. 2008,
  The Astrophysical Journal, 674, L33

\bibitem[{Hosokawa \& Omukai(2009)}]{Hosokawa09}
Hosokawa, T. \& Omukai, K. 2009, The Astrophysical Journal, 691, 823

\bibitem[{Howard {et~al.}(1997)Howard, Koerner, \& Pipher}]{Howard97}
Howard, E.~M., Koerner, D.~W., \& Pipher, J.~L. 1997, Astrophysical Journal,
  477, 738

\bibitem[{Keto(2002)}]{Keto02}
Keto, E. 2002, The Astrophysical Journal, 568, 754

\bibitem[{Keto(2003)}]{Keto03}
Keto, E. 2003, The Astrophysical Journal, 599, 1196

\bibitem[{Keto \& Wood(2006)}]{KW06}
Keto, E. \& Wood, K. 2006, The Astrophysical Journal, 637, 850

\bibitem[{Klaassen {et~al.}(2012)Klaassen, Testi, \& Beuther}]{K12}
Klaassen, P.~D., Testi, L., \& Beuther, H. 2012, Astronomy and Astrophysics,
  538, A140

\bibitem[{Klaassen \& Wilson(2007)}]{KW07}
Klaassen, P.~D. \& Wilson, C.~D. 2007, The Astrophysical Journal, 663, 1092

\bibitem[{Klaassen \& Wilson(2008)}]{KW08}
Klaassen, P.~D. \& Wilson, C.~D. 2008, The Astrophysical Journal, 684, 1273

\bibitem[{Klaassen {et~al.}(2011)Klaassen, Wilson, Keto, Zhang,
  Galv{\'a}n-Madrid, \& Liu}]{K11}
Klaassen, P.~D., Wilson, C.~D., Keto, E.~R., {et~al.} 2011, Astronomy and
  Astrophysics, 530, 53

\bibitem[{Kurtz {et~al.}(1994)Kurtz, Churchwell, \& Wood}]{Kurtz94}
Kurtz, S., Churchwell, E., \& Wood, D. O.~S. 1994, The Astrophysical Journal
  Supplement Series, 91, 659

\bibitem[{L{\'o}pez-Sepulcre {et~al.}(2009)L{\'o}pez-Sepulcre, Codella,
  Cesaroni, Marcelino, \& Walmsley}]{LS09}
L{\'o}pez-Sepulcre, A., Codella, C., Cesaroni, R., Marcelino, N., \& Walmsley,
  C.~M. 2009, Astronomy and Astrophysics, 499, 811

\bibitem[{Mezger \& Henderson(1967)}]{mezger67}
Mezger, P.~G. \& Henderson, A.~P. 1967, Astrophysical Journal, 147, 471

\bibitem[{Mottram {et~al.}(2011)Mottram, Hoare, Davies, Lumsden, Oudmaijer,
  Urquhart, Moore, Cooper, \& Stead}]{Mottram11}
Mottram, J.~C., Hoare, M.~G., Davies, B., {et~al.} 2011, The Astrophysical
  Journal, 730, L33

\bibitem[{Muzerolle {et~al.}(1998)Muzerolle, Hartmann, \& Calvet}]{Muzerolle98}
Muzerolle, J., Hartmann, L., \& Calvet, N. 1998, The Astronomical Journal, 116,
  2965

\bibitem[{Okamoto {et~al.}(2003)Okamoto, Kataza, Yamashita, Miyata, Sako,
  Takubo, Honda, \& Onaka}]{okamoto03}
Okamoto, Y.~K., Kataza, H., Yamashita, T., {et~al.} 2003, The Astrophysical
  Journal, 584, 368

\bibitem[{Peters {et~al.}(2010)Peters, Banerjee, Klessen, Low,
  Galv{\'a}n-Madrid, \& Keto}]{Peters:2010bw}
Peters, T., Banerjee, R., Klessen, R.~S., {et~al.} 2010, The Astrophysical
  Journal, 711, 1017

\bibitem[{Peters {et~al.}(2012)Peters, Klaassen, Mac~Low, Klessen, \&
  Banerjee}]{with_thomas_12}
  Peters, T., Klaassen, P.~D., Mac~Low, M.-M., Klessen, R.~S., \& Banerjee, R.
  2012, The Astrophysical Journal, 760, 91
  
  \bibitem[{Wei {et~al.}(2008)Wei, Woody, Teuben, La~Vigne, \& Vogel}]{wei08}
Wei, L., Woody, D., Teuben, P., La~Vigne, M., \& Vogel, S. 2008,
  {http://www.mmarray.org/memos/carma\_memo45.pdf}

\bibitem[{Wilson {et~al.}(2009)Wilson, Rohlfs, \& H{\"u}ttemeister}]{Tools5}
Wilson, T.~L., Rohlfs, K., \& H{\"u}ttemeister, S. 2009, Astronomy and
  Astrophysics Library, Vol.~5, {Tools of Radio Astronomy} (Berlin, Heidelberg:
  Springer Berlin Heidelberg)

\bibitem[{Wu {et~al.}(2004)Wu, Wei, Zhao, Shi, Yu, Qin, \& Huang}]{Wu04}
Wu, Y., Wei, Y., Zhao, M., {et~al.} 2004, Astronomy and Astrophysics, 426, 503

\end{thebibliography}

\appendix

\section{HCO$^+$ Channel Maps}
\begin{figure}[htb!]
\centering
\includegraphics[width=\textwidth]{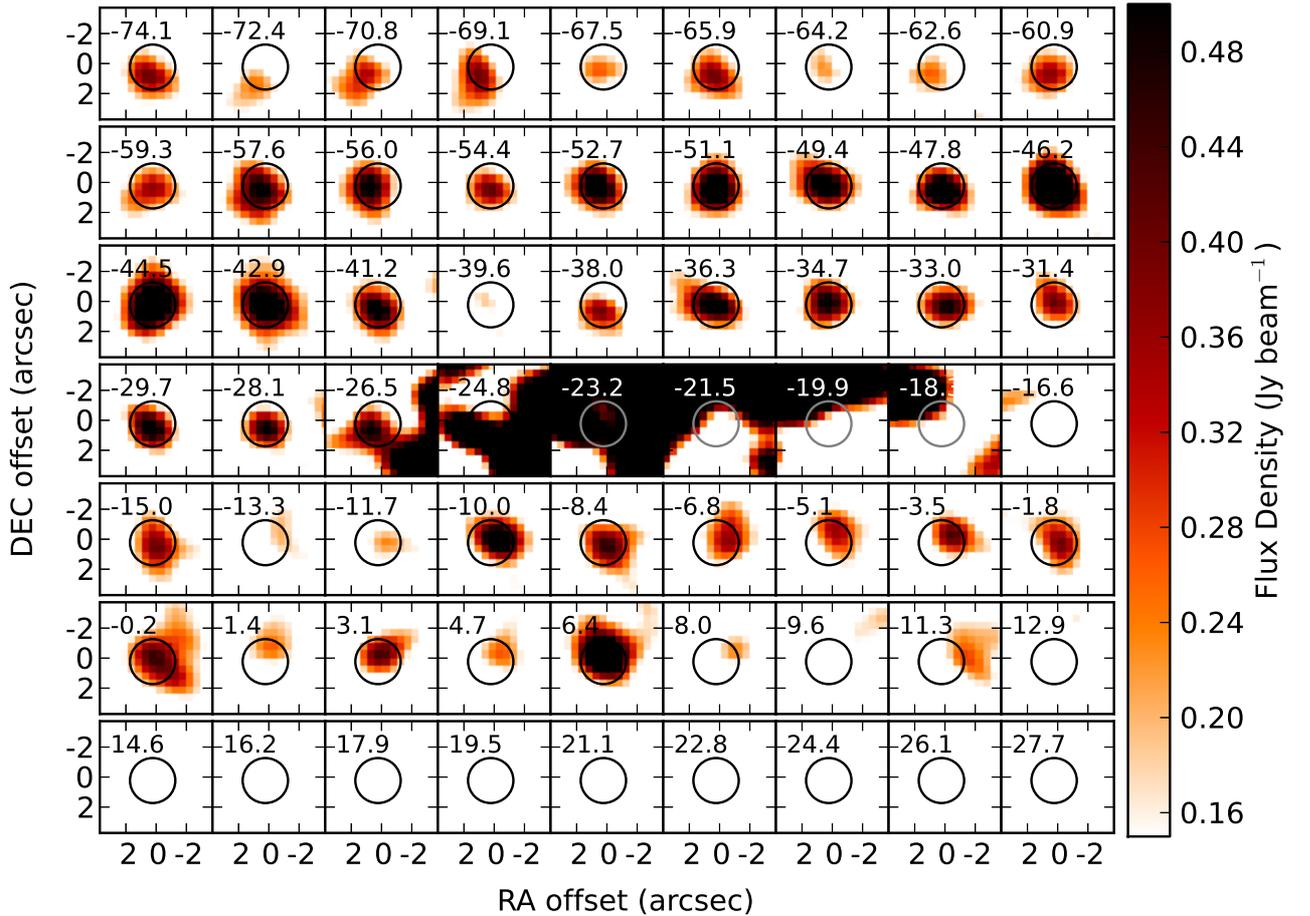}
\caption{Channel map of small scale outflowing HCO$^+$.  The black circle here is the same as the white circle in the HCO$^+$ first moment map of Figure \ref{fig:outflow}. The number in the top left of each panel shows the central velocity (in km s$^{-1}$) of the channel being shown.}
\label{fig:channel_map}
\end{figure}
Here we show the channel map of the HCO$^+$ J=1-0 emission centered on the entrained molecular outflow described in the main text.  The black circle at the center of each channel map is the same circle as shown in the HCO$^+$ first moment map of Figure \ref{fig:outflow}. As can be seen in the central velocity channels, the emission from the envelope material overwhelms the outflow emission.  At higher blue and red shifted velocities, the outflowing material becomes much more prominent.

\acknowledgements
The authors would like to thank the editor for the effort he put into this manuscript.  T.P. acknowledges financial support through SNF grant 200020\_137896. R.G.-M. and S.N.L acknowledge
funding from the European Community's Seventh Framework Programme (/FP7/2007-2013/) under grant agreement No 229517

\end{document}